\documentclass[preprint]{elsarticle}%    

\usepackage{graphicx}
\usepackage{bm}
\usepackage{epsf}
\usepackage{rotating}
\usepackage{epsfig,graphics,rotate,color}
\usepackage{amssymb}
\usepackage{amsmath}
\usepackage{amsfonts}
\usepackage{array,hhline,dcolumn}%
\providecommand{\U}[1]{\protect\rule{.1in}{.1in}}
\begin{document}
\title{A Study of the Fluorescence Response of Tetraphenyl-butadiene}
\author{R. Jerry}
\address{Physics Dept., Howard University, Washington, DC 20059}
\cortext[cor1]{Corresponding author:  L. Winslow, lwinslow@mit.edu}

\author{L. Winslow, L. Bugel and  J.M. Conrad}
\address{Physics Dept., Massachusetts Institute of Technology,
Cambridge, MA 02139}

\begin{abstract}
  Tetraphenyl-butadiene (TPB) is a widely used fluorescent
  wavelength-shifter.  A common application is in liquid-argon-based
  particle detectors, where scintillation light is produced in the
  UV at 128 nm.  In liquid argon experiments, TPB is often employed to
  shift the scintillation light to the visible range in order to allow
  detection via standard photomultiplier tubes.  This paper presents
  studies on the stability of TPB with time under exposure to light.
  We also examine batch-to-batch variations.  We compare
  scintillation-grade TPB to 99\% pure TPB response.  In the 99\% pure
  samples, we report a yellowing effect, and full degradation of the
  TPB emission-peak, upon extended exposure to light.

\end{abstract}
\maketitle

\section{Introduction}

1,1,4,4-Tetraphenyl-1,3-butadiene (TPB) is used as a wavelength
shifter in many experiments.  For example, TPB is used in numerous
liquid argon (LAr) experiments including ICARUS \cite{ICARUS}, WARP
\cite{WARP}, MiniCLEAN \cite{clean,deapclean}, DEAP
\cite{deapclean,DEAP} and MicroBooNE \cite{uBooNE} to shift
scintillation light, which is produced at 128 nm in LAr, to the
visible, where it can be detected by photomultiplier tubes (PMTs).
TPB absorbs light across a broad range from the high UV to its
emission peak, which lies at approximately $425\pm50$ nm.

TPB has the chemical formula (C$_6$H$_5$)$_2$C=CHCH=C(C$_6$H$_5$)$_2$.
Scintillation-grade TPB, which is used in LAr experiments, has an
assay of $>99\%$ purity.  TPB dissolves in toluene, and this mixture is
often used when coating surfaces such as PMTs.  One can spray
or paint the TPB-toluene mixture onto the surface and allow the toluene to
evaporate, leaving only a layer of TPB behind.  Alternatively, one
can dissolve polystyrene into the mix to produce a
thin plastic skin with embedded TPB when the toluene evaporates.

In this paper, we investigate whether the TPB fluorescence response
degrades over time when the chemical is exposed to light.  This is
relevant to the handling of the TPB-coated materials as experiments
are assembled.

\section{Methods}

All studies used samples of 0.25 g of TPB dissolved in 100 ml of
toluene.  We found that tests using dissolved mixtures in a cuvette,
as opposed to tests of TPB powder in a cuvette, gave a more repeatable
measured response when the sample was removed and replaced in the
test-stand spectrometer

We studied two grades of TPB: scintillation grade TPB (SG-TPB) and
99\% pure TPB.  For consistency in the studies over time, all samples
of a given grade were taken from the same batch of TPB powder
purchased form Sigma-Aldrich \cite{SA}.  In our studies of freshly
mixed samples, we also report on results of TPB powder purchased from
ACROS \cite{ACROS}. All TPB powder batches were kept in opaque
brown bottles in a dark cabinet throughout the study.

For studies of exposure to light, TPB-toluene mixtures were stored in
clear glass bottles.  These were illuminated by a SOLUX solar spectrum
lamp \cite{SOLUX} for periods of three months and one month.  The
samples were maintained at room temperature during the exposure to the
SOLUX light.

\begin{figure}\begin{center}
{
\includegraphics[height=3.1in]{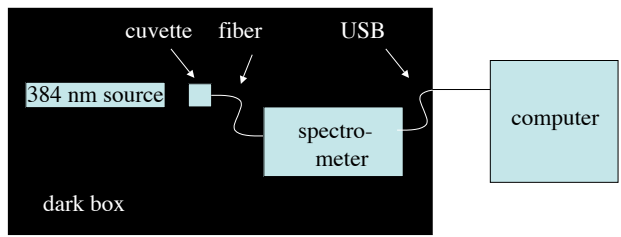}}\end{center}
\vspace{-1.in}
\caption{\label{setup}  A diagram of the setup for the measurements.
The light source illuminates the cuvette.  The light may be transmitted
or absorbed and re-emitted at a longer wavelength.  The spectrometer
measures the wavelength of light which exits the back of the cuvette.}
\end{figure}

The experimental setup used to measure the fluorescence of the
TPB-toluene samples is shown in Fig.~\ref{setup}.  The samples were
held in a cuvette with (1 cm)$^2$ cross section, which was illuminated
by a 384-nm light source. This particular wavelength transmits through
toluene, which absorbs UV light with $<$315 nm
wavelength\cite{KayeLaby}.  On the other hand, 384 nm is within the
absorption range of TPB.  The fluorescence spectrum from the samples
is piped to a spectrometer via a fiber coupled to the back of the
cuvette.   We use a StellarNet Black Comet (model C) spectrometer
\cite{SpectraWiz}.  
This is a
concave grating spectrometer with a range from 190 to 850 nm.

\begin{figure}[t]\begin{center}{
\includegraphics[height=3.1in]{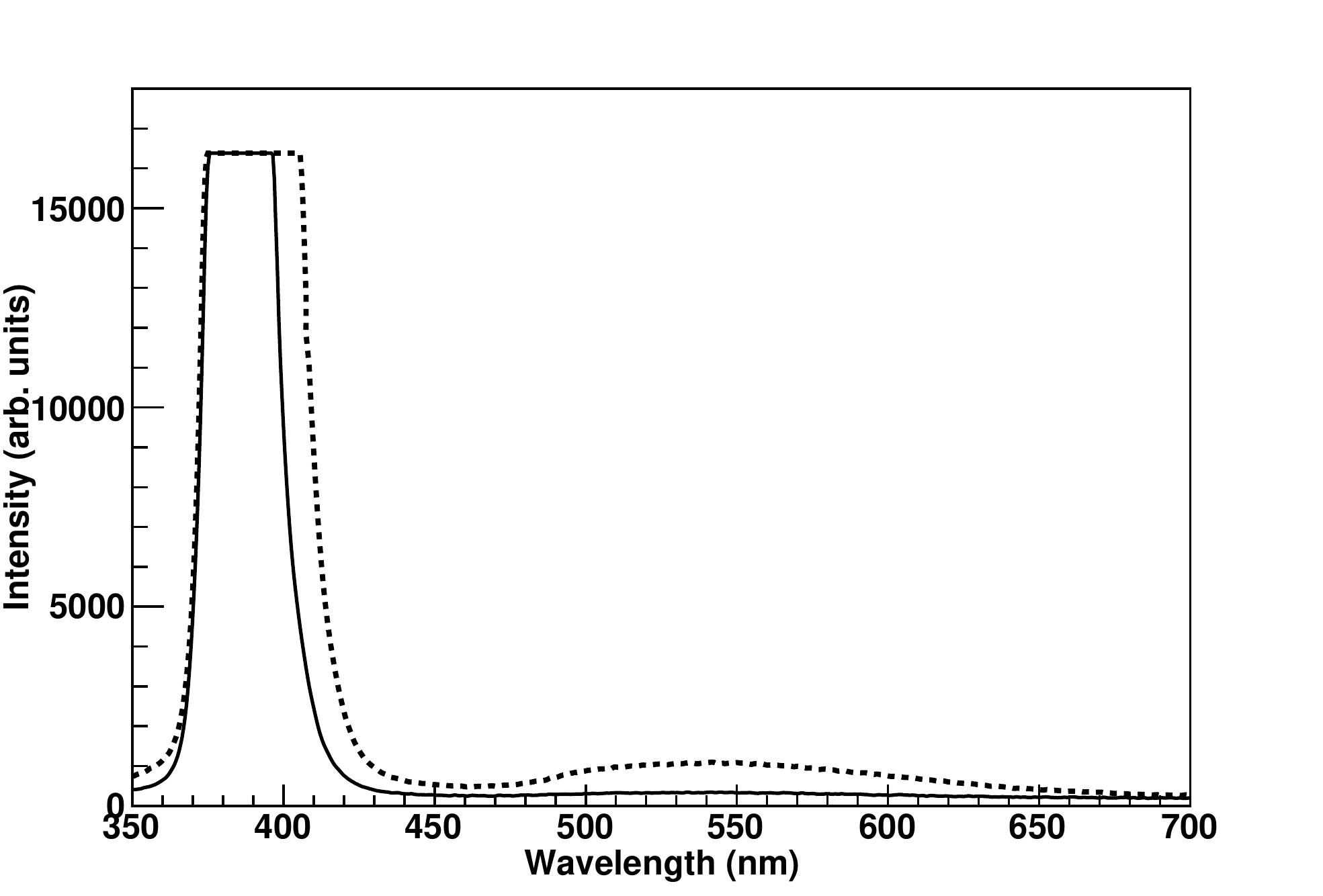}
}\end{center}
\caption{\label{base}  Solid:  The light source as observed through the empty cuvette.  Dashed: The light source observed through a cuvette filled with toluene.   The broad toluene peak, centered at about 550 nm can be seen.  The peak from 
the light source, which is at 384 nm,  is truncated so that the toluene peak is visible.}
\end{figure}

Fig.~\ref{base} (solid) shows the peak of the 384 nm light source as
observed through the empty spectrometer cuvette, while Fig.~\ref{base}
(dashed) shows the spectrometer measurement when the cuvette is filled
with pure toluene.  One can see that the toluene produces a broad peak
centered at about 550 nm.  The stability of the height of the toluene
peak provides a useful benchmark for our study.  We do not expect the
toluene fluorescence peak to substantially change with exposure to
light, and so this peak provides a method to normalize the absolute
light measurements, which can vary with minor changes of the alignment
of the light source to the spectrometer.

\begin{figure}\begin{center}
{
\includegraphics[height=3.1in]{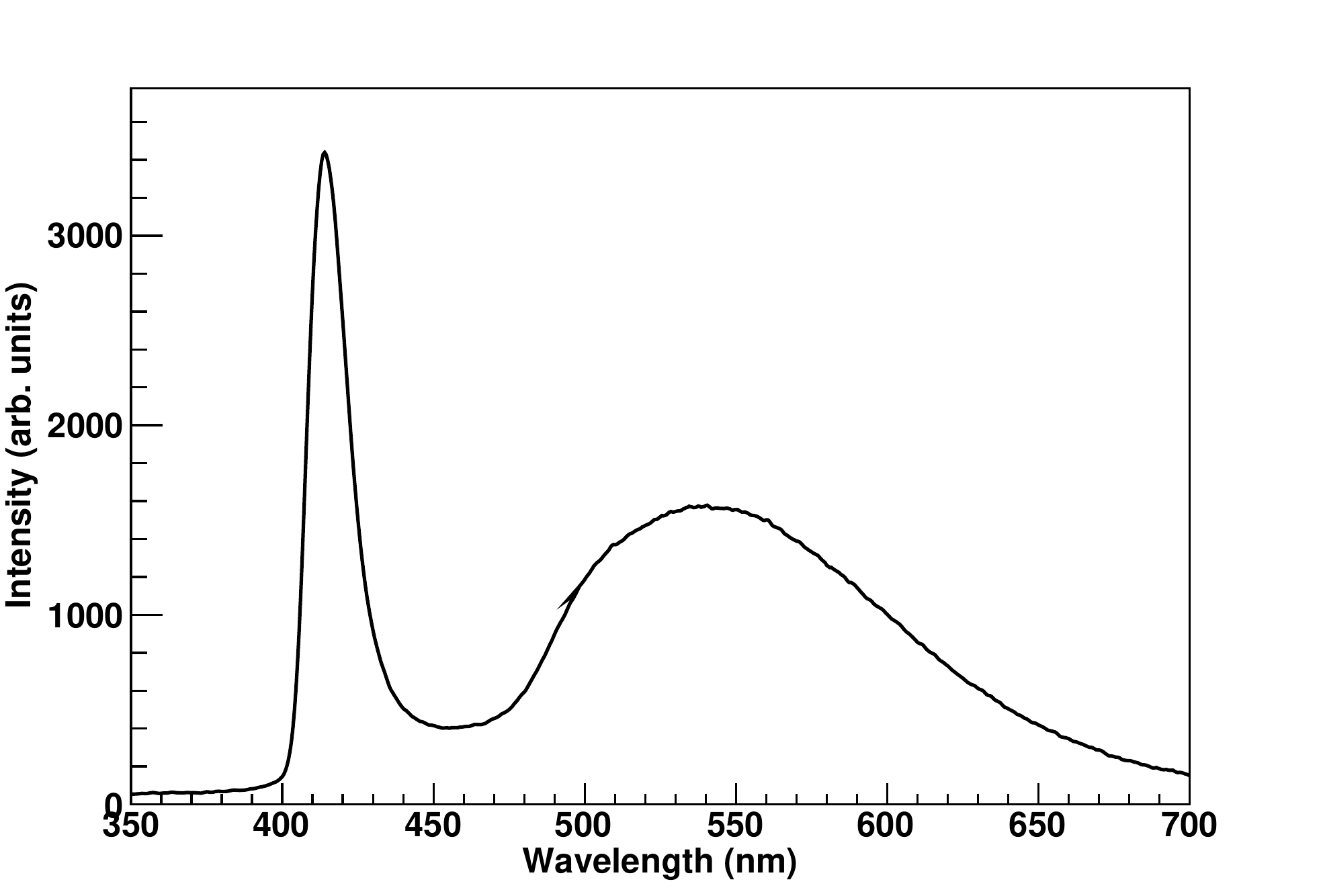}}\end{center}
\caption{\label{immediate} Spectrum from freshly-mixed TPB and 
toluene.  The TPB peak and toluene peak are visible, while
  the peak from the light source is fully absorbed.}
\end{figure}

Fig.~\ref{immediate} shows the resulting spectrum from a cuvette
filled with freshly-mixed SG-TPB--toluene solution.  The light from
the source is absorbed by the TPB, and so there is no visible peak at
384 nm.  One can see a strong peak at 425 nm from the TPB and the
broad peak from the toluene.  This curve provides a basis of
comparison for other samples in this study.

When normalizing to the toluene peak at 550 nm, we observed that the
peaks of freshly-mixed samples of the same TPB batch vary from
lowest to highest counts by 10\%.  We present this below as a $\pm
5\%$ systematic error on the measurement.  We believe this is
consistent with the expected combined error of the measurement of the
mass of TPB powder and the toluene volume during mixing.

\begin{figure}\begin{center}
{\vspace{-1.in}
\includegraphics[height=3.75in]{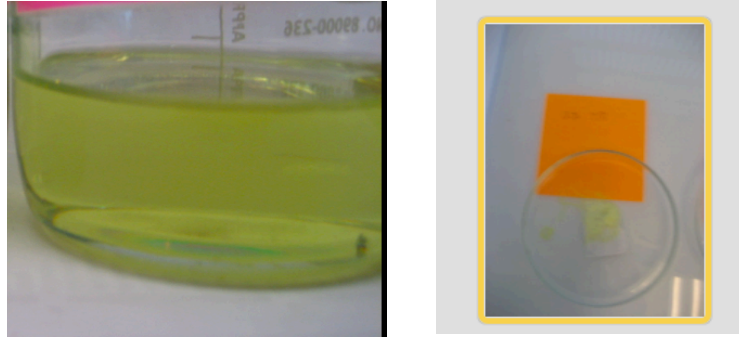} \vspace{-1.in}}\end{center}
\caption{\label{yellow} Left:   Toluene and 99\% TPB mixture exposed to light for
three days.  The initially clear mixture became yellow-green.   Right:  Powdered 99\%
pure TPB exposed to the SOLUX lamp for five days.  The initially white powder became yellow.}
\end{figure}

As noted above, we also provide a comparison of the SG-TPB response to
99\% pure TPB, which is generally considered poor for use as a
wavelength shifter, but which is also less expensive.  During this
phase of the study, we observed a striking effect.  When a sample of
99\% pure TPB was mixed with toluene and exposed to ambient room
light, within the time-scale of a few days, the originally clear
mixture became yellow-green, as shown in Fig.~\ref{yellow} (left).  We
also noticed yellowing when the 99\% pure TPB powder was left exposed
to ambient room light for three days, as shown in Fig.~\ref{yellow}
(right).  This effect can be reproduced with the SOLUX full-spectrum
light source.  We report on this further below.

\section{Results}

{\begin{figure}\begin{center}
{
\includegraphics[height=3.1in]{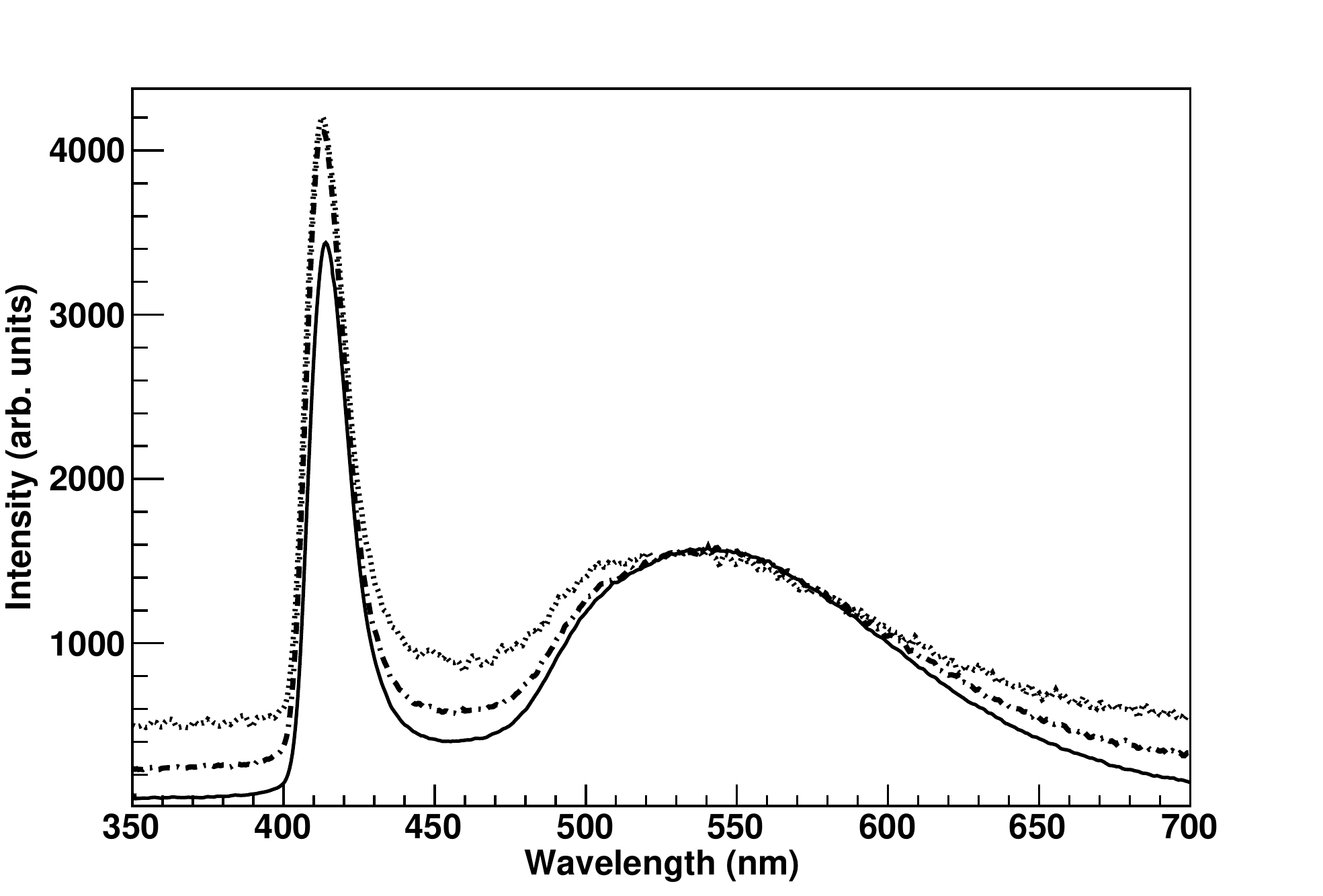}
~\\
\includegraphics[height=3.1in]{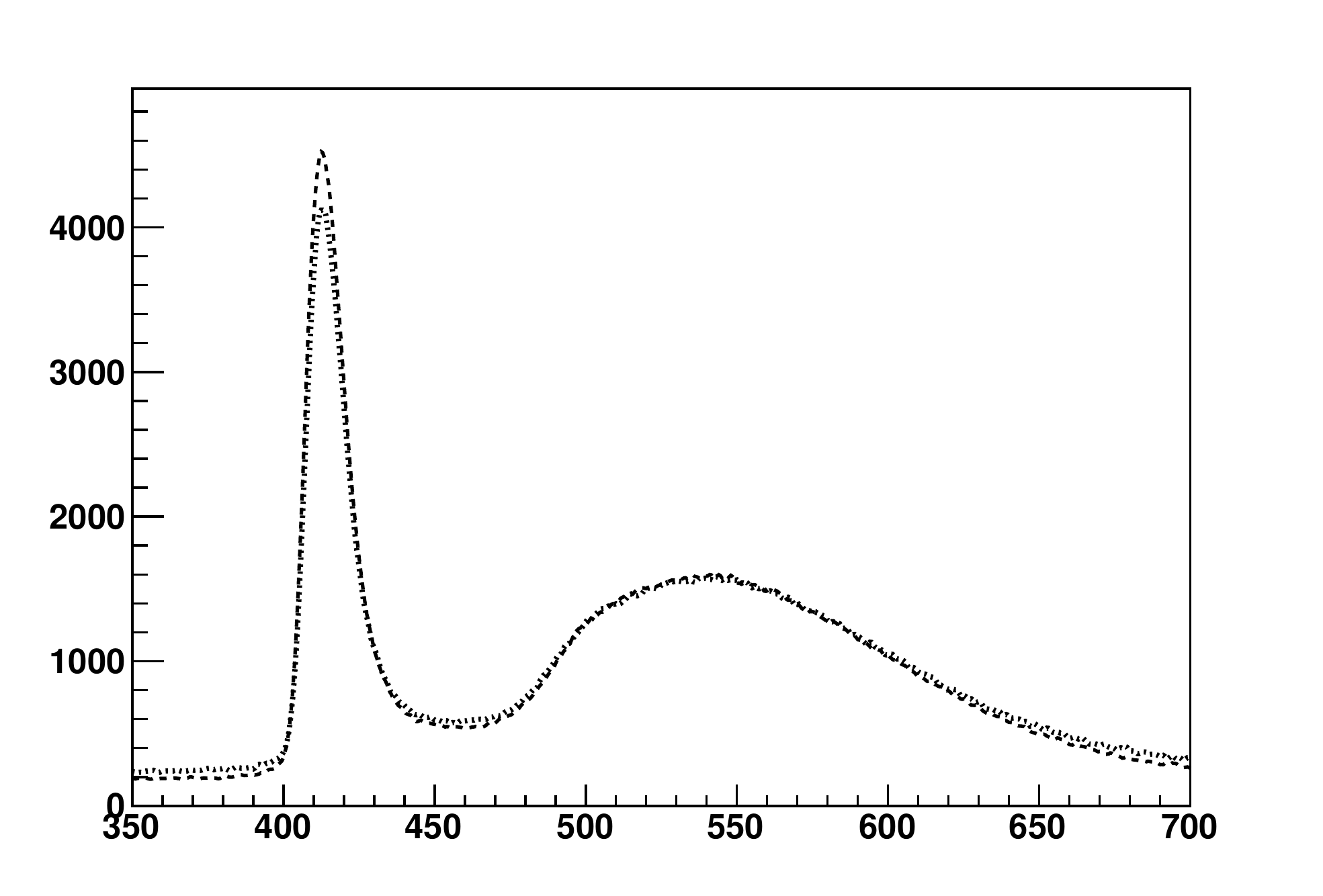}}\end{center}
\caption{\label{sg} Top: Spectra for scintillation grade TPB dissolved
  in toluene for samples exposed to SOLUX light source for three
  months (dot-dashed) and one month (dotted) compared to the spectrum
  from a freshly mixed sample (solid). Bottom: Spectrum for
  scintillation grade TPB dissolved in toluene exposed to SOLUX lamp
  for three months (dotted) compared to a sample stored in the dark
  for three months (dashed).  In both sets of plots, there is agreement
  within the systematic error of this study.}
\end{figure}}

\subsection{Scintillation Grade TPB Exposed to Light 
and Over Time}

Fig.~\ref{sg} (top) compares the results of SG-TPB--toluene
mixtures exposed to the light as a function of time.  For reference,
the solid line is the freshly-mixed SG-TPB--toluene.  The
dashed curve shows the response of the mixture exposed for one month,
while the dot-dashed line shows the mixture exposed for three months.
The height of the toluene peak at 550 nm was used for relative
normalization of the curves.  Within the $\pm5\%$ systematic error, the TPB peak
responses are in agreement.

We kept one SG-TPB--toluene mixed sample for a period of three months
in a dark cabinet.  In Fig.~\ref{sg} (bottom), we compare the transmitted
light from this sample (dashed line) to the freshly-mixed (solid
line) and the mixture with three month exposure to the SOLUX light source (dotted line).
Again, the TPB peaks are in agreement.

\subsection{Comparison of Three Batches of Scintillation Grade TPB}

\begin{figure}\begin{center}
{
\includegraphics[height=3.1in]{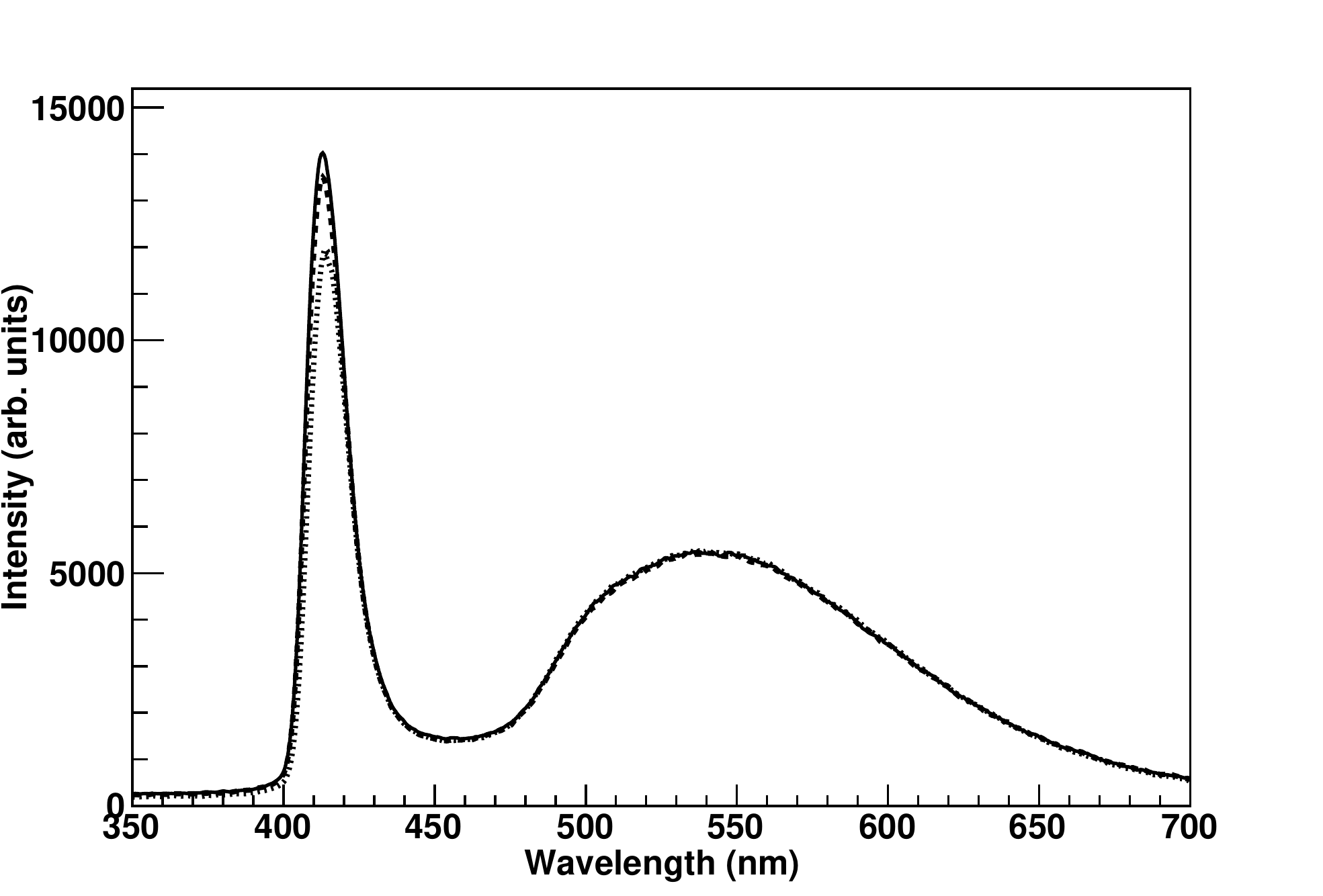}}\end{center}
\caption{\label{allsg} Spectra for scintillation grade TPB dissolved
  in toluene, freshly mixed.  Results from three separate batches are
  compared.  Solid: batch recently purchased from Sigma Aldrich;
  Dotted: batch purchased from Sigma-Aldrich six months earlier;
  Dashed: batch purchased recently from ACROS.  All results are in
  good agreement within the $\pm5\%$ systematic error.}
\end{figure}

The initial batch of TPB from Sigma-Aldrich was compared to a batch
from Sigma-Aldrich purchased six months later and to a batch purchased
from ACROS \cite{ACROS}.  The results of this cross comparison is shown in
Fig.~\ref{allsg}.  One can see that the samples agree well within the
$\pm5\%$ systematic error.

\subsection{Comparison of Scintillation Grade and 99\% Pure TPB Response}

\begin{figure}\begin{center}
{
\includegraphics[height=3.1in]{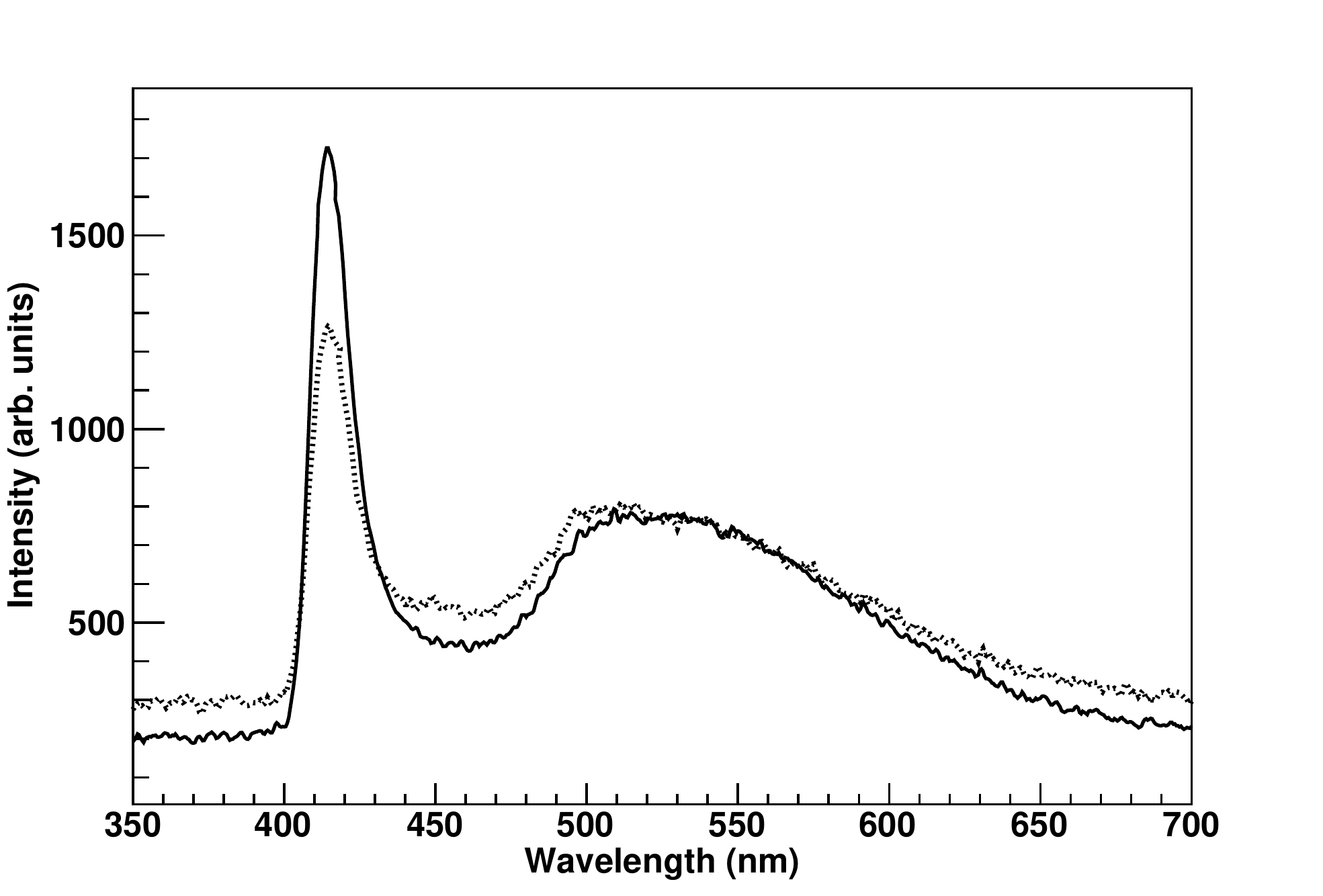}\\
\includegraphics[height=3.1in]{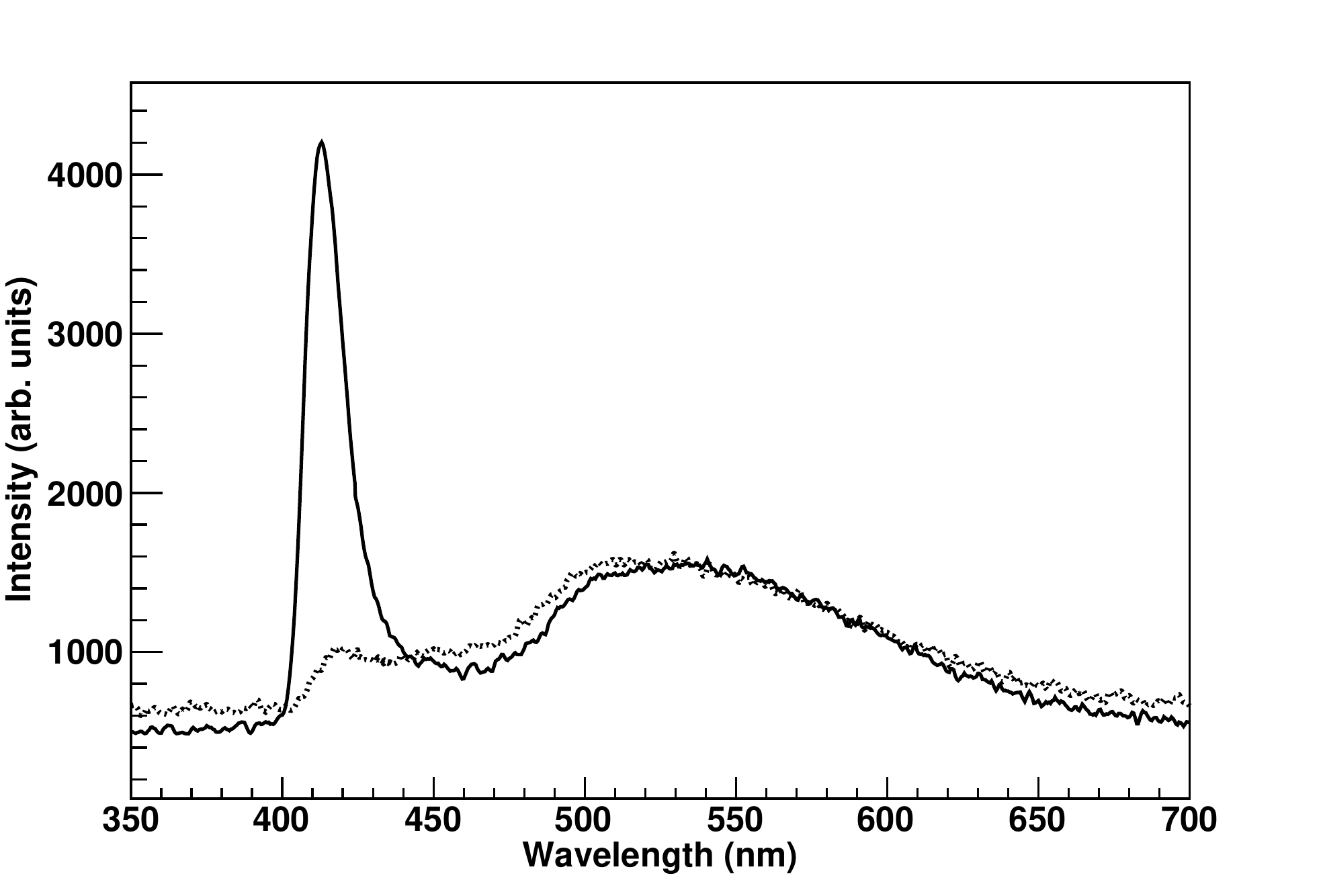}}\end{center}
\caption{\label{sg99} Spectra for scintillation grade TPB dissolved in
  toluene (solid) compared to 99\% pure TPB dissolved in toluene
  (dotted).  Top: Freshly mixed samples. The 99\% pure TPB response is
  about two thirds of the level of the scintillation grade.  Bottom:
  After exposure to light where the 99\% pure TPB mixture has shown
  yellowing.  The TPB peak is substantially reduced in the yellowed
  mixture. }
\end{figure}

Next, we compare the response of SG-TPB and 99\% pure TPB mixtures in
Fig.~\ref{sg99} (top) solid and dashed lines, respectively.   From this,
we conclude that the light output from 99\% pure TPB is reduced 
compared to SG-TPB, as expected.

We observed that the 99\% pure TPB-toluene mixture suffered yellowing
with exposure to sunlight, ambient room light, and also to the SOLUX
lamp light.  This effect was observed in three separately acquired
batches of 99\% pure TPB ordered months apart. We did not see this
effect with any batch of SG-TPB.  Recently, we learned that the WARP
experiment has seen a similar yellowing of one batch of SG-TPB mixed
in toluene \cite{flavio}.

In our studies, the yellowing was visible after about three days of
exposure to light and did not visibly change after this, over a period
of one month.  The spectrum from the yellowed mixture is compared to
the one-month exposure of SG-TPB in Fig.~\ref{sg99} (bottom).  The TPB
peak of the yellowed mixture is nearly entirely absorbed.

In order to demonstrate that the yellowing is due to exposure to
light, a 99\% pure TPB-toluene mixture was produced and divided.  Half
half exposed to light immediately and half was stored in a dark
cabinet.  After a period of about three days, the mixture in the light
showed clear yellowing while the mixture stored in the dark looked
clear.  The latter mixture was removed from the cupboard and set into
the light.  After about three days of exposure, yellowing appeared.

\section{Conclusions}

We have investigated the fluorescence response of TPB dissolved in
toluene.  We studied the effect of exposure to light for extended
periods and conclude that within the level of $\pm5\%$ systematic
variations, the scintillation grade TPB is stable to exposure to
light.  This would indicate that TPB-coated equipment does not require
special protection from light.

We compared three batches of scintillation-grade TPB and found that the
relative response agreed within the systematic error.  We also
showed that scintillation-grade TPB has superior response
compared to 99\% pure TPB.

The 99\% pure TPB showed yellowing upon exposure to light over a
period of several days.  At this point, we do not know the cause of
the yellowing.  It appears to be an impurity in the TPB which reacts
when exposed to light.  While we have not observed yellowing in the
three scintillation-grade batches we studied, there was a report from
WARP of one scintillation-grade batch which showed yellowing which may
be the same effect. Because the initial mixture appears clear, with
yellowing occurring over several days, we recommend that all batches
of TPB, even those which are scintillation grade, be tested for this
effect before use.  A simple test is to place a small amount of the
TPB powder from each batch into a petri dish and expose it to light
for one week before use.  Batches which show yellowing should be
rejected.

\section*{Acknowledgments}
The authors thank B. Jones for producing samples for study of the
yellowing of 99\% pure TPB.  We thank the MicroBooNE PMT working group
for useful discussion.  We thank the Guggenheim Foundation, the
National Science Foundation, and the MIT Summer Research Program for
support.

\end{document}